\documentstyle[aps,manuscript]{revtex}
\input{epsf}

\begin{document}
\def\sech{\mathop{\rm sech}\nolimits}

\title{Generation and manipulation of squeezed states of
light in optical networks for quantum communication and computation}

\author{Maxim Raginsky and Prem Kumar}

\address{Center for Photonic Communication and Computing \\
Department of Electrical and Computer Engineering \\
Northwestern University, Evanston, Illinois 60208-3118}

\maketitle

\begin{abstract}

We analyze a fiber-optic component which could find multiple uses in novel
information-processing systems utilizing squeezed states of light.   Our approach is based
on the phenomenon of photon-number squeezing of soliton noise after
the soliton has propagated through a nonlinear optical fiber.
Applications of this component in optical networks for quantum computation and quantum cryptography
are discussed.

\end{abstract}

\newpage

Information-processing systems where information is carried by nonclassical
states of light \cite{yuen} (e.g., photon number states or squeezed states),
though not as yet implemented with standard telecommunications in
mind, nevertheless present an attractive alternative for such novel
applications as quantum computation \cite{gottesman} or secure
quantum-key distribution \cite{gottesman1}. As
envisioned currently \cite{braunstein,gottesman}, such systems can be
constructed using linear optics, linear mixing of creation and annihilation
operators (linear Bogoliubov transformations which include
squeezing as a special case), and nonlinear operations for state
preparation and detection. 

One major obstacle to current practical implementation of such systems is the
lack of components which would preserve the noise characteristics of
the transmitted information.  This requirement is crucial because squeezed states are highly
sensitive to loss, as a simple analysis reveals.  In optical communication networks, optical amplifiers
are inserted along the fiber links in order to restore exponentially decaying
signal power to acceptable levels.  These devices, represented by an
idealized model of a phase-insensitive amplifier (PIA), add at least 3
dB of noise with coherent-state inputs when the amplifier gain is high
\cite{dariano} (this minimal noise figure is referred as the Standard Quantum Limit, or
SQL \cite{sqlnote}) and thus may destroy precisely engineered noise statistics of the
transmitted states.  The quantum-key distribution scheme of Gottesman
and Preskill \cite{gottesman1}, for example, calls for at
least 2.51 dB of squeezing in channels with weak noise, so introducing
a PIA into the underlying communications infrastructure may severely
affect the performance of the scheme.

In this Letter we concentrate on a component of such
information-pocessing systems that could prove useful for generating and manipulating squeezed states of light, photon-number squeezed (PNS) states in particular.  As recent experimental reports indicate \cite {experiments},
it is possible to generate sub-Poissonian light (a near-PNS state) by means
of the following simple setup.  A soliton pulse is launched into an
optical fiber and then frequency-filtered at the fiber exit.  The resulting
quantum-mechanical state of light exhibits photon-number fluctuations
below the coherent-state level, with maximum squeezing observed when the
fiber is three soliton periods long.  To illustrate the utility of
this setup to our goal, we analyze the arrangement in which the above
process is iterated by launching the filtered soliton through another
fiber-and-filter stage. From here on, we will refer to our setup as
the Dual-Stage Squeezer (DSS), as opposed to the Single-Stage Squeezer
(SSS) described at the beginning of the paragraph.

Our motivation to suggest the use of the DSS in
optical networks for quantum information processing comes
from the observation, further elaborated below, that the DSS can be
thought of as the original SSS operating on
squeezed (rather than simply coherent) input states. This property of the
DSS makes it a natural choice for easily implementable sources of
squeezed states, whose degree of squeezing can be controlled by tuning
the soliton-pulse parameters, and also for devices that enhance the
degree of squeezing introduced into the quantum network. For reasons mentioned above, such squeezing enhancers would
conceivably play a key role in practical implementations of the
Gottesman-Preskill quantum key distribution scheme using squeezed
states. \cite{gottesman1}

In a recent study, Levandovsky {\em et al.} \cite{levandovsky} have
employed the soliton perturbation approach \cite{haus} to obtain a
complete theoretical description of the quantum-noise statistics of
spectrally filtered solitons. This linearization approach is valid
whenever the photon-number noise is small compared to the average
number of photons in the soliton, which is usually the case in most
experiments.  We brifely recount the main idea of
their approach and then apply it to our analysis of the DSS.

Nonlinear evolution of an electromagnetic pulse
propagating through a lossless optical fiber is
governed by the quantum nonlinear Schr\"odinger equation
\begin{equation}
\frac {\partial} {\partial \xi} \hat{a} (\tau,\xi) = i \left[ \frac
{1} {2}
 \frac {\partial^2} {\partial \tau^2} + \hat{a}^\dag (\tau,\xi)
\hat{a} (\tau,\xi)
\right] \hat{a} (\tau,\xi),
\label{qnlse}
\end{equation}
where $\hat{a}(\tau,\xi)$ is the annihilation operator of the field and
$(\xi,\tau)$ are the dimensionless space and time coordinates.  The
corresponding classical equation has a fundamental soliton solution
$a_0(\tau,\xi)={\rm e}^{i \xi / 2} \sech \tau \equiv f_0(\tau) {\rm
e}^{i \xi / 2}$, given here in canonical form with two photons per pulse.  We write the annihilation
operator as
\begin{equation}
\hat{a}(\tau,\xi) = [f_0(\tau)+\Delta\hat{a}(\tau,\xi)]{\rm
e}^{i \xi/2},
\label{perturb}
\end{equation}
where $\Delta\hat{a}$ is the annihilation operator that represents the
perturbation of the soliton mean field by quantum noise and satisfies the usual equal-space commutation relations,
$[\Delta\hat{a}(\tau,\xi),\Delta\hat{a}(\tau^{\prime},\xi)]=[\Delta\hat{a}^\dag(\tau,\xi),\Delta\hat{a}^\dag(\tau^{\prime},\xi)]=0$,
$[\Delta\hat{a}(\tau,\xi),\Delta\hat{a}^\dag(\tau^{\prime},\xi)]=\delta(\tau-\tau^{\prime})$,
everywhere inside the fiber. We make the linearizaiton approximation
by substituting Eq.~(\ref{perturb}) into Eq.~(\ref{qnlse}) and discarding
all terms that are $O (\Delta \hat{a}^2)$, thus
separating the problem into the classical NLSE for the mean field $a_0
(\tau,\xi)$ and the linearized operator equation
\begin{equation}
\frac {\partial \Delta \hat{b}} {\partial \xi} = \frac {i} {2} \frac
{\partial^2} {\partial \tau^2} \Delta \hat{b} + 2i |a_0 (\tau,\xi)|^2
\Delta \hat{b} + i a^2_0 (\tau,\xi) \Delta \hat{b}^\dag
\label{lin_qnlse}
\end{equation}
for $\Delta\hat{b} = \Delta\hat{a} {\rm e}^{i \xi/2}$. In what follows, we shall
disregard phase factors of the form ${\rm e}^{i \theta \xi}$, $-
\infty < \theta < \infty$, because the filtered light is
directly detected in a SSS.

The solution of Eq.~(\ref{lin_qnlse}) can be written as an eigenfunction expansion \cite{haus}
\begin{equation}
\Delta \hat{a} (\tau,\xi) = \frac {1} {2 \pi} \int [\hat{V}_c
(\Omega,\xi) f_c (\Omega,\tau) + \hat{V}_s (\Omega,\xi) f_s
(\Omega,\tau)] d\Omega  + \sum_{i=n,p,\tau,\theta} \hat{V}_i (\xi) f_i(\tau)
\label{perturb_expand}
\end{equation}
with operator coefficients, where the discrete eigenmodes $f_n$, $f_p$,
$f_\tau$, and $f_\theta$ represent perturbations of the soliton mean
field due to changes in photon number, momentum, time, and phase
respectively; and $f_c$ and $f_s$ are the symmetric and anti-symmetric
continuum eigenmodes that represent perturbation of the dispersive
radiation in the fiber.  Detailed analysis of these modes, along
with their time-domain and frequency-domain forms, can be found in
Ref.~\onlinecite{haus}.

The $\xi$-dependent Hermitian operators $\hat{V}_i$, $i
\in M \equiv \{
c,s,n,p,\tau,\theta \}$, are obtained by projecting
Eq.~(\ref{perturb_expand}) onto the eigenmodes $\{ \tilde {f}_i\,|\,i \in
M \}$ of the equation adjoint
to Eq.~(\ref{lin_qnlse}) which is obtained by reversing the sign of
the $\Delta \hat{b}^\dag$ term.  The relevant orthogonality relations
are $\langle f_i , \tilde{f}_j \rangle = \Delta_{ij}$, where the inner
product is defined by $\langle f, \tilde{g} \rangle \equiv {\rm Re}
\int f(\tau) \tilde{g}^* (\tau) d \tau$; $\Delta_{ij}$ =
$\delta_{ij}$ in all cases except for $i=j \in \{ c,s \}$, where
$\Delta_{ij} = 2 \pi \delta (\Omega - \Omega^{\prime})$.

Defining the time-domain cosine quadrature operator $\Delta \hat{a}_c = (\Delta \hat{a} + \Delta
\hat{a}^\dag)/2$, we write the time-domain correlation
function $G(\tau,\tau' ; \xi) = 4 \langle \Delta \hat{a}_c (\tau,\xi) \Delta
\hat{a}_c (\tau^{\prime},\xi) \rangle$ and the corresponding
covariance function $C(\tau,\tau^{\prime} ; \xi) = G(\tau,\tau^{\prime};\xi) - 4
\langle \Delta \hat{a}_c (\tau,\xi) \rangle \langle \Delta \hat{a}_c
(\tau^{\prime}, \xi) \rangle$.  Assuming that the filter $H(\omega)$
at the fiber exit is linear and imposing the realizability condition
$0 \le |H(\omega)| \le 1$, we obtain the perturbation operator after
the filter from the frequency-domain relation
\begin{equation}
\Delta\hat{a}_{\rm out}(\omega,\xi)=|H(\omega)| \Delta
\hat{a}(\omega,\xi) + \sqrt{1 - |H(\omega)|^2} \hat{v}(\omega),
\end{equation}
where $\hat{v}$ is a vacuum-state operator associated with the
frequency-dependent loss due to the filter.  The observed squeezing $S(\xi)$ is quantified
by normalizing the output photon-number variance to the average output
photon number:
\begin{equation}
S(\xi) = 1 + \frac {1} {4 \pi^2 \langle \hat{N}_{\rm out} \rangle}
\int \int d\omega d\omega^{\prime} f_0(\omega) |H(\omega)|^2 C_N(\omega,\omega^{\prime};\xi)
|H(\omega^{\prime})|^2 f_0(\omega^{\prime}),
\label{squeezing}
\end{equation}
where $C_N(\omega,\omega^{\prime};\xi) = C(\omega,\omega^{\prime};\xi)
- 2 \pi \delta (\omega - \omega^{\prime})$ is the normally ordered
part of the Fourier transform $C(\omega,\omega^{\prime};\xi)$ of the
covariance function $C(\tau,\tau^{\prime};\xi)$, and $f_0(\omega) = \pi
\sech \frac {\pi \omega} {2}$ is the Fourier transform of $f_0(\tau)$.

As stated by Levandovsky {\em et al.} \cite{levandovsky}, in the case of the SSS the correlation function $G(\omega,\omega^{\prime};\xi)$ is equal to the covariance
function $C(\omega,\omega^{\prime};\xi)$ because the perturbation at the fiber entrance ($\xi =
0$) is white coherent-state quantum noise.  The observed squeezing in
dB, given by $-10 \log S(\xi)$, is shown in Fig.~\ref{stage1} for
the case of a bandlimited parabolic filter \cite{mecozzi} $H(\omega) = 1 - \omega^2 /
\eta^2$, $|\omega| \le \eta$, where the bandwidth $\eta$ is adjusted
to give 10\% loss.  The analytical expression for the correlation
function $G$ in time domain, as well as the observed squeezing for
other types of filters, are given in Ref.~\onlinecite{levandovsky}.

Now we turn to the analysis of the DSS.  In order to render the problem
tractable while retaining its essential physical features, we have made the following assumptions: (a) the filter in
the first stage is weak enough, so that the quantum-mechanical average
of its output may be treated as
the soliton mean field $a_0(\tau,\xi)$ plus a small perturbation $\Delta b_0(\tau,\xi) \equiv \Delta a_0(\tau) {\rm e}^{i
\xi/2}$; (b) all relevant frequency-domain quantities are narrowband \cite{narrowband_note},
so that the number of photons at high frequencies that are cut off
with a bandlimited filter is negligible, and the mean field can still
be treated as a fundamental soliton; and (c) as the pulse propagates
through the second stage, the quantum-mechanical average $\langle \Delta \hat{a}(\tau) \rangle$
remains independent of $\xi$ and is equal to $\Delta a_0 (\tau)$. This last
assumption can be given precise mathematical meaning by ensuring that
the maximum steady-state $(\tau \rightarrow \infty$) error which
results from assuming that $\Delta a_0(\tau) {\rm e}^{i \xi/2}$ is a
solution of the linearized Eq.~(\ref{lin_qnlse}) is vanishingly
small \cite{error_note}.  With these assumptions in place, the DSS becomes equivalent to
the SSS, but with a squeezed-state input and an additional filter inserted
between the soliton source and the entrance to the fiber.

Since the filter is assumed to be weak, we may write its transfer
function in the form $H(\omega) = 1 + h(\omega)$, where $h(\omega)$ is
small, so that $\Delta a_0 (\omega) = h(\omega) f_0(\omega)$ is the
frequency-domain form of the quantum-mechanical average $\langle
\Delta \hat{a}(\tau) \rangle$ which,
as we have assumed, is independent of $\xi$. Therefore, at the fiber
entrance to the second stage, the soliton mean field is perturbed by an
ideal squeezed state with the quantum-mechanical average $\Delta
a_0(\tau)$ and the squeezing parameter $r$ determined by the fiber
length and the filter transfer function of the first stage. It is then
easy to see that the time-domain covariance function
for the DSS is given by
\begin{equation}
C_{\rm DSS}(\tau,\tau^{\prime};\xi) = {\rm e}^{-2
r} G_{\rm SSS}(\tau,\tau^{\prime};\xi) - 4 \Delta a_0 (\tau) \Delta a_0
(\tau^{\prime}),
\label{cov_dss}
\end{equation}
where $G_{\rm SSS}(\tau,\tau^{\prime};\xi)$ is the correlation function for the
SSS and $\xi = 0$ at the fiber entrance to the second stage.  The observed squeezing $S_{\rm DSS}(\xi)$ at the output of
the DSS is then
computed by substituting the normally ordered covariance function
$C_{{\rm DSS},N}(\omega,\omega^{\prime};\xi)$ into
Eq.~(\ref{squeezing}).

We have analyzed numerically the following arrangement.  If the fiber
in the first stage is three soliton periods long, then with our
particular filter the maximum squeezing is 2.8 dB, which corresponds
to $r \simeq 0.32$. The initial perturbation $\Delta a_0 (\omega)$ is given by $- \frac {\omega^2} {\eta^2} f_0 (\omega)$ (for all
frequencies $|\omega| \le \eta$, a condition which is sufficient in accordance
with our narrowband assumption).  The maximally squeezed output of the
SSS is then launched through the second stage.  The observed squeezing
at the output of the DSS with the first stage generating maximally
squeezed output is shown in Fig.~\ref{stage2} as a function of the
fiber length in the second stage, with maximum squeezing of 6.1 dB
observed if the fiber in the second stage is three solition periods
long.

The results of the preceding analysis suggest that by iterating the SSS it is
possible to construct a wide variety of components for generation and manipulation of squeezed states in optical networks for
quantum communication and computation.  As already mentioned, the SSS
can be used to generate near-PNS states whose statistical parameters
can be tuned by varying the pulse power and width. The main incentive
to use the DSS for generation of near-PNS states is the
simplicity of implementation: in order to get more squeezing, we just
add another stage to the SSS with another parabolic filter, thus
avoding the use of optimized filters whose shapes are complicated and
difficult to implement. \cite{levandovsky}  In other words, compared to a SSS, one gets more
squeezing out of a DSS while using less fiber and easily realizable
filters.

As for manipulation of squeezed states in such networks, the DSS, with
more than a twofold increase in squeezing (as measured in decibels), renders an
excellent illustration of how one can use fiber nonlinearity to enhance the
noise statistics of light transmitted through the network. Such an enhancement
would be needed, e.g. if the squeezing introduced previously was degraded by
loss. To summarize, the capabilities offered by fiber nonlinearity need to be
closely explored with such novel applications as quantum information
processing in mind.
\\
\\
The authors acknowledge useful discussions with D. Levandovsky. This
work was supported in part by the U.S. Army Research Office through
MURI grant DAAD19-00-1-0177.

\newpage

\begin{figure}
\caption{Observed squeezing (in dB) vs. fiber length (in soliton
periods) in the SSS.  The
filter has a bandlimited parabolic frequency response giving 10\% loss.}
\label{stage1}
\end{figure}

\begin{figure}
\caption{Observed squeezing (in dB) vs. second-stage fiber length (in soliton
periods) in the DSS with the first stage generating maximally squeezed
output (the first-stage fiber is 3 soliton periods long).  The filter
has a bandlimited parabolic frequency response giving 10\% loss.}
\label{stage2}
\end{figure}

\newpage

\vspace*{\fill}
\centerline{\epsfbox{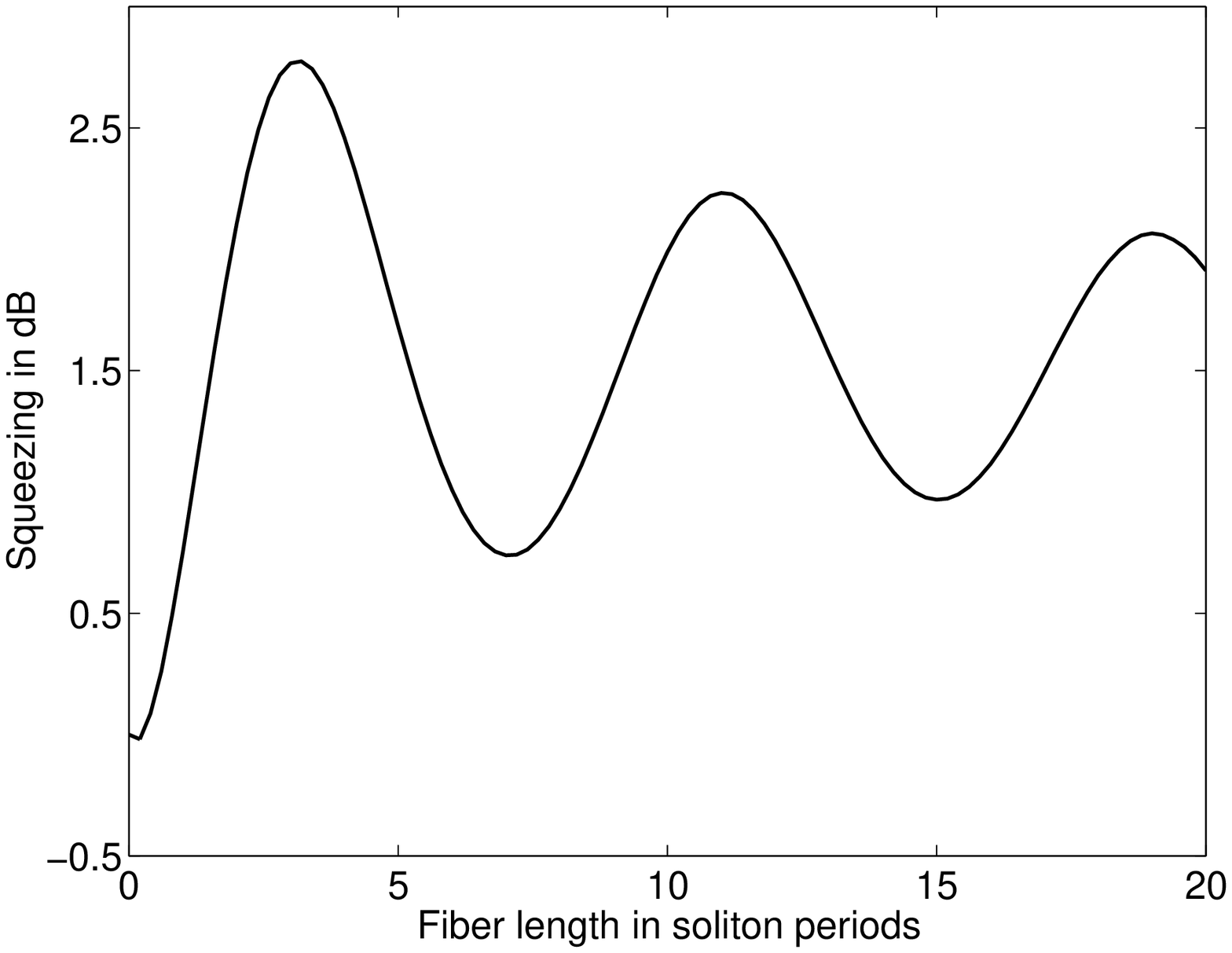}}
\vspace*{\fill}

Fig.~\ref{stage1} -- ``Generation and manipulation \ldots''
by Raginsky and Kumar.

\newpage

\vspace*{\fill}
\centerline{\epsfbox{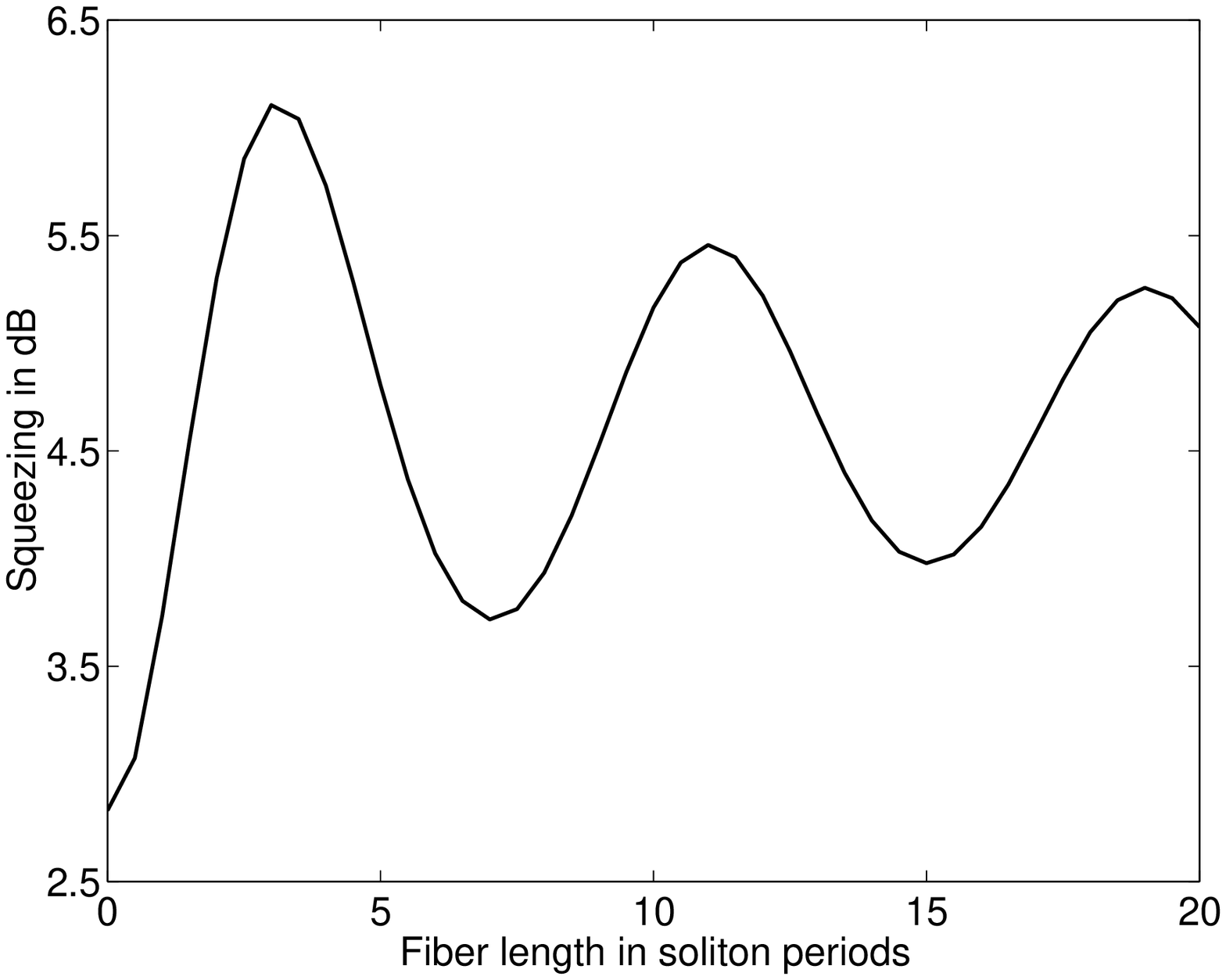}}
\vspace*{\fill}

Fig.~\ref{stage2} -- ``Generation and manipulation \ldots''
by Raginsky and Kumar.


\begin{references}
\bibitem{sqlnote} The term ``Standard Quantum Limit'' (SQL) comes up
in the literature on quantum measurements (e.g., position monitoring of a free mass) and is usually taken to
signify some sort of a fundamental bound imposed by the
quantum-mechanical nature of the problem at hand.  In the context of
quantum amplifiers, the SQL refers to the ultimate lower bound on the
noise figure of a high-gain PIA with coherent-state inputs.
\bibitem{narrowband_note}The assumption that all the relevant
frequency-domain quantities are narrowband (i.e. have bandwidths small
compared to the baseband frequency) is equivalent to the time-domain
slowly varying envelope approximation.  We invoke this assumption in
order to make the numerics more efficient.  A more thorough treatment
would necessarily call for determining the exact form of the
post-filter emergent soliton.
\bibitem{error_note}The steady-state error is given by
\begin{displaymath}
\epsilon_{\rm steady} = \lim_{\tau \rightarrow \infty} \sup_\xi \left| 
\frac {\partial f(\tau,\xi)} {\partial \xi} -
{\mathcal L}[f(\tau,\xi)]\right|,
\end{displaymath}
where $f(\tau,\xi)$ is the approximate solution of the $c$-number
version of Eq.~(\ref{lin_qnlse}) and ${\mathcal L}[\cdot]$ is the
linear operator acting on the (mean) perturbation on the right-hand side of
Eq.~(\ref{lin_qnlse}).  In our case, $f(\tau,\xi)
=(1/\eta^2)\exp(i\xi/2)d^2 \sech\tau/d\tau^2$, and the error is given
by
\begin{displaymath}
\epsilon_{\rm steady}=\lim_{\tau \rightarrow 0} \frac{6 \sech^3 \tau
\tanh^2 \tau}{\eta^2}=0.
\end{displaymath}
In fact, the quantitiy under the limit sign rolls off very rapidly in $\tau$.
\end{references}
\end{document}